\documentstyle[epsfig]{aipproc}

\begin{document}
\title{Broad-band continuum and variability of NGC 5548}

\author{
Pawe\l\ Magdziarz$^1$, 
Omer~Blaes$^2$,
Andrzej~A.~Zdziarski$^3$,
W.~Neil~Johnson$^4$ and
David~A.~Smith$^5$
}
\address{
$^1$Astronomical Observatory, Jagiellonian University, 
Orla 171, 30-244 Cracow, Poland
\\
$^2$Department of Physics, University of California, 
Santa Barbara, CA 93106
\\
$^3$N. Copernicus Astronomical Center, 
Bartycka 18, 00-716 Warsaw, Poland
\\
$^4$Naval Research Lab., Code 7651, 
4555 Overlook Ave., SW, Washington, DC 20375-5352
\\
$^5$Department of Physics and Astronomy, University of Leicester, 
Leicester, LE1 7RH, UK
}

\maketitle

\begin{abstract}

We analyze a composite broad-band optical/UV/X$\gamma$-ray spectrum of
the Seyfert 1 galaxy NGC 5548. The spectrum consists of an average of
simultaneous optical/{\it IUE}/{\it Ginga\/} observations accompanied by
{\it ROSAT\/} and {\it GRO}/OSSE data from non-simultaneous observations.
We show that the broad-band continuum is inconsistent with simple disk models
extending to the soft X-rays. Instead, the soft-excess is well described by
optically thick, low temperature, thermal Comptonization which may
dominate the entire big blue bump. This might explain the observed tight
UV/soft X-ray variability correlation and absence of a Lyman edge
in this object. However, the plasma parameters inferred by the spectrum
need stratification in optical depth and/or temperature to prevent
physical inconsistency. The optical/UV/soft X-ray component contributes
about half of the total source flux. The spectral variations of the
soft-excess are consistent with that of the UV and argue that the
components are closely related. The overall pattern of spectral
variability suggests variations of the source geometry, and shows the
optical/UV/soft X-ray component to be harder when brighter, while the
hard X-ray component is softer when brighter. 

\end{abstract}

\section*{Introduction}

NGC~5548 is one of the brightest Seyfert 1 galaxies with
optical/UV/X$\gamma$ continuum consisting of three characteristic
components: the big blue bump, the soft-excess, and the hard X-ray thermal
Comptonization continuum with the reflection hump. These components are
known to be highly variable, but there have been no observing campaigns
covering all of them simultaneously. The only simultaneous {\it IUE}/{\it
Ginga} campaign (Clavel et al.\ 1992; Nandra et al.\ 1993) showed a tight
correlation between the UV bump and hard X-ray component, thus supporting
UV/X-ray reprocessing models. Recently, Marshall et al.\ (1997) have found
a correlation between the UV and the soft X-ray component which suggests
that the soft-excess also participates in the reprocessing. We
analyze a composite broad-band optical/UV/X$\gamma$ spectrum in order to
develop a continuum model and apply it to re-analyze the variability
correlations in a simultaneous sub-set of available {\it IUE}/{\it Ginga}
observations. 

\begin{figure}[b!] 
\centerline{\epsfig{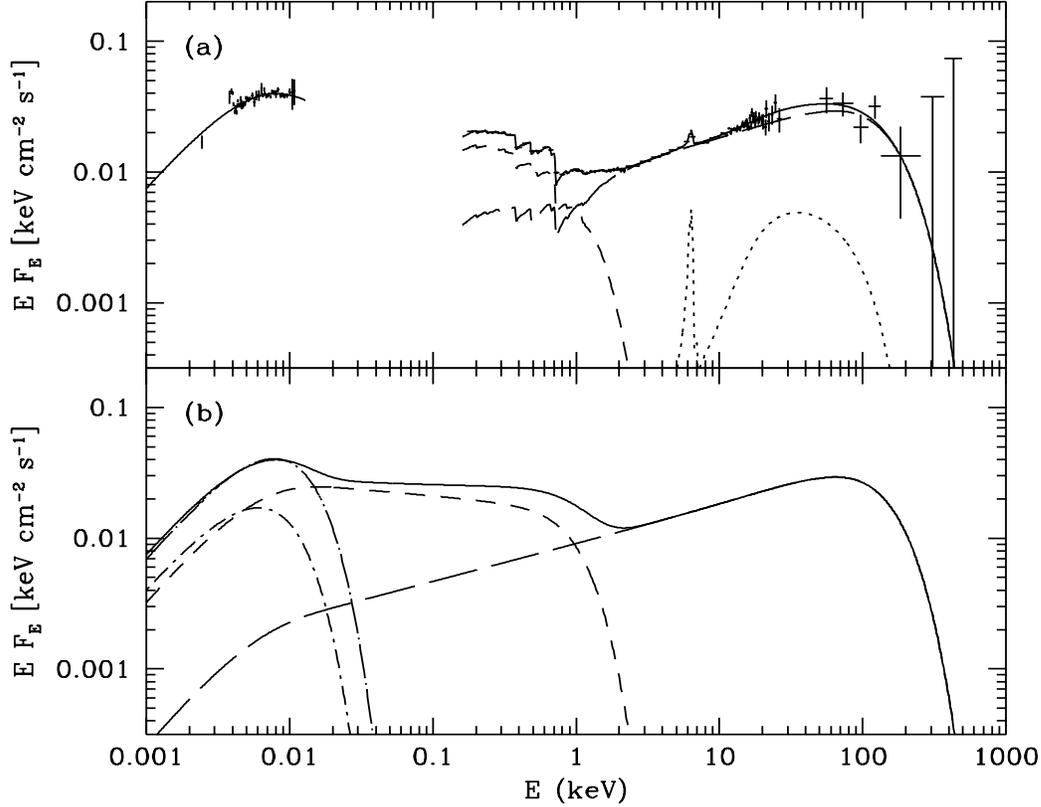}}
\vspace{10pt}
\caption{
The intrinsic (i.e., after the effect of Galactic absorption at the soft
excess and optical/UV reddening have been removed) broad-band composite
optical/UV/X$\gamma$ spectrum of NGC~5548. The top panel shows the data and
fitted model. The spectrum consists of an average over the simultaneous
sub-set of
optical/{\it IUE}/{\it Ginga} data supplemented with non-simultaneous
average {\it ROSAT} and {\it GRO}/OSSE observations. The non simultaneous
data were fitted with free normalizations and renormalized on the figure.
The bottom panel shows a deconvolution of the spectrum into continuum
components. The dot-short-dashed and the short-dashed curves show the disk
and the soft-excess component respectively. The long-dash and the dotted
(in top panel)
curves show the thermal Comptonization coronal spectrum and the reflection
component respectively. The solid curve gives the resulting model. The
dot-long-dash curve shows the disk component fitted to the optical/{\it
IUE} data alone. The energy scale is in the source frame.
} 
\label{fig1} 
\end{figure}

\section*{Broad-band Spectrum and Variability}
 
The optical/{\it IUE} and {\it Ginga} data come from the simultaneous sub set
of the Jan.~1989--July 1990 campaign reported by Clavel et al.\ (1992) and
Nandra et al.\ (1991) respectively. The {\it ROSAT} data is an average
spectrum over a monitoring campaign of Dec.~1992--Jan.~1993 (Done et al.\
1995). The {\it GRO}/OSSE data were compiled by McNaron-Brown et al.\ (1997, in
preparation) from Phase 1 and 3 observations. We use the {\it Ginga} data
from both the top-layer and mid-layer of the LAC detector. The mid-layer
gives more effective area in the 10--20 keV range which is crucial for
determining the Compton reflection component. That layer had previously been
ignored due to problems with background subtraction. 

\subsection*{ Broad-band Continuum }

We find the average spectra from non-simultaneous observations to be
consistent
within statistical discrepancies with that from the simultaneous
campaign.  However, the results have to be treated carefully since the
averaging of the highly variable spectrum produces very wide confidence limits. 
The soft-excess component contributes significantly up to an energy $\sim$2
keV (Fig.~1a) which is marginally observed in the {\it Ginga} data. As
was shown by Fiore, Matt, and Nicastro (1997), this component can not be
simply explained by atomic processes.  The {\it GRO}/OSSE spectra are
consistent with a constant high energy cut-off at about 100 keV, and no
variations of the spectral shape.  This gives Comptonizing plasma
parameters of $kT_{HC}\sim$50 keV and $\tau\sim$2, consistent with
those typical
of Seyfert 1s (Zdziarski et al.\ 1997). Our re-analysis of the {\it
ROSAT} campaign shows that the average data well constrain the
soft-excess spectral index, $\Gamma$=$2.1^{+0.3}_{-0.2}$, but do not
constrain the energy cut-off.  This, however, is well established at
E$_{SE}$=$0.6\pm 0.1$ keV from {\it Ginga} observations. The average {\it
IUE} data well constrain both the reddening of the spectrum
E(B$-$V)=$0.03\pm 0.01$, and the maximum disk temperature $kT_d$=$3.2\pm
0.2$ eV. This argues against the high temperature disk (in agreement with 
Laor et al.\
1997). 

\begin{figure}[b!] 
\centerline{\epsfig{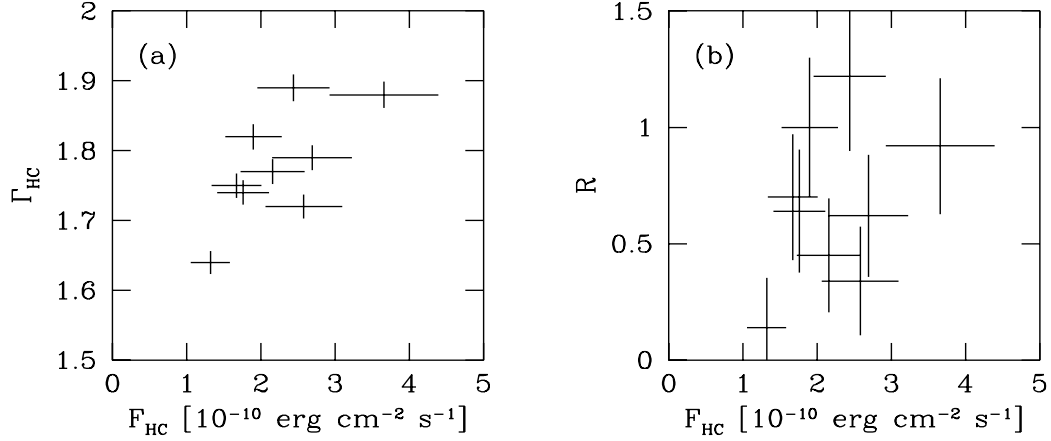}}
\vspace{10pt}
\caption{
Correlation between the total flux emitted in the X/$\gamma$ hard
continuum with (a) photon spectral index, and (b) amount of reflection 
($\Omega/2\pi$).
When the source is brighter, it shows a softer spectral index and larger solid
angle of cold matter intercepting the hard continuum.
}
\label{fig2}
\end{figure}

\begin{figure}[b!] 
\centerline{\epsfig{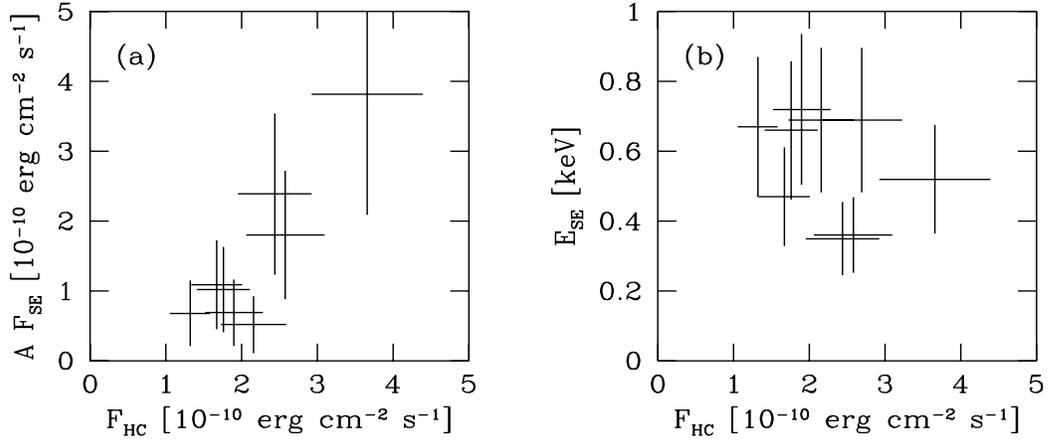}}
\vspace{10pt}
\caption{
Correlation of (a) the total flux emitted in the soft-excess, and (b) the
cut-off energy of the soft-excess component with the total flux emitted in
the X/$\gamma$ hard continuum. The total flux emitted in the soft excess
is positively correlated with that from the hard continuum, while the
cut-off energy of the soft excess remains constant. The soft excess is
modeled by a cut-off power-law with the spectral index frozen at
$\Gamma$=2.2 (cf. Walter \& Fink 1993). The plotted F$_{SE}$ flux is
the soft excess flux multiplied by a model dependent constant, A. 
}
\label{fig3}
\end{figure}

It is energetically possible for the entire blue bump (the dot-long-dash
curve in Fig.~1b) to arise from reprocessing of the hard continuum
assuming the X/$\gamma$ source forms a patchy corona above the surface of
the disk. However, the nature of the soft excess remains unclear in a such
model and it is hard to explain the tight correlation in simultaneous {\it
IUE}--{\it EUVE} observations (Marshall et al.\ 1997). On the other hand,
extrapolation of the {\it ROSAT} soft X-ray power-law points exactly to
the UV component, suggesting a Comptonization origin. Then the blue bump
turns out to be dominated by Comptonization (the short-dash, and the
dot-short-dash in Fig.~1b), and contains about half the total flux. 
However, Comptonizing the disk component by a hypothetical cold plasma
requires an extreme optical depth ($\tau$$\sim$30), 
which needs to be explained by
stratification of temperature and optical depth (Nandra et al.\ 1995). 

\subsection*{ Variability Correlations }

The hard X-ray continuum shows significant correlation between the total flux
changes and the spectral index and amount of reflection (Fig.~2a, b).
Such a pattern of spectral variability suggests changes in the source
geometry which produce more reflection when the source is brighter and
softer. The total X/$\gamma$ flux is correlated with the total optical/UV
flux (cf.\ Nandra et al.\ 1993), but the energy balance is dependent
on the spectral model assumed. 
The soft excess component required by the {\it Ginga} data seems to vary
in a correlated fashion with the total flux emitted by the X/$\gamma$
continuum (Fig.~3a). The cut-off energy of the soft-excess is consistent
with being constant over the hard continuum changes (Fig.~3b). The above
results suggest that the soft-excess component is related to
reprocessing. Marshall et al.'s (1997)  results
suggest much higher variability amplitude in the EUV than in the UV.
Hence one should
expect variability in the soft-excess spectral index, which is consistent
with the overall `harder when brighter' variability pattern in the optical/UV,
but opposite to the `softer when brighter' pattern in the hard X/$\gamma$
continuum.

\section*{Conclusions}

We have shown that the soft-excess component in NGC~5548 may contribute to the
energy reprocessing and its variability behavior is consistent with that of the
big blue bump. The soft-excess is consistent with optically
thick Comptonization, probably in a complex, stratified
plasma. If the soft-excess is actually related to optically thick
Comptonization, the entire big blue bump and the soft-excess may be
dominated by the same component. This would explain both the tight {\it
IUE}--{\it EUV} correlation (Marshall et al.\ 1997), and the absence of a Lyman
edge in the UV spectrum (Kriss et al.\ 1997).

\end{document}